\documentclass{svmult}

\usepackage{makeidx}         
\usepackage{graphicx}        
\usepackage{multicol}        
\usepackage[bottom]{footmisc}

\makeindex             

\begin{document}

\title*{From Network Structure to Dynamics and Back Again:
Relating dynamical stability and
connection topology in biological complex systems}
\titlerunning{From Network Structure to Dynamics and Back Again} 
\author{Sitabhra Sinha}
\institute{The Institute of Mathematical Sciences, CIT Campus, Taramani,
Chennai 600113, India
\texttt{sitabhra@imsc.res.in}}
%
%
\maketitle

The recent discovery of universal principles underlying many complex 
networks occurring across a wide range of length scales in the biological
world has spurred physicists in
trying to understand such features using techniques from statistical physics 
and non-linear dynamics. In this paper, we look at a few examples of 
biological networks to see how similar questions can come up in very
different contexts. We review some of our recent work 
that looks at how network structure (e.g., its connection topology) 
can dictate the nature of its dynamics, and conversely, how dynamical 
considerations constrain the network structure. We also see how
networks occurring in nature can evolve to modular configurations as a result
of simultaneously trying to satisfy multiple structural and dynamical 
constraints. The resulting optimal networks possess hubs and have heterogeneous
degree distribution similar to those seen in biological systems.

\section{Introduction}
\begin{quote}
To see a world in a grain of sand,\\
And a heaven in a wild flower,\\
Hold infinity in the palm of your hand,\\
And eternity in an hour.\\
~~~~~~~~-- William Blake, {\em Auguries of Innocence}
\end{quote}
\label{sec:1}

Like Blake, physicists look for universal principles that are valid across 
many different systems, often spanning several length or time scales.
While the domain of physical systems has often offered examples of such 
widely applicable `laws', biological phenomena tended to be,
until quite recently, less fertile in terms of generating similar 
universalities, with the notable exception of allometric scaling 
relations~\cite{SN84}.
However, this situation has changed with the study of complex networks 
emerging into prominence. Such systems comprise a large number of 
nodes (or elements) linked with each other according to specific connection 
topologies, and are seen to occur widely across the biological, 
social and technological worlds~\cite{AB02,Newman03,DM03}. Examples
range from the intra-cellular signaling system that consist
of different kinds of molecules affecting each other via
enzymatic reactions, to the internet composed of servers
around the world which exchange enormous quantities of 
information packets regularly, and food webs which link,
via trophic relations, large numbers of inter-dependent
species. While the existence of complex networks in various
domains had been known for some time, the recent
excitement among physicists working on such systems has
to do with the discovery of certain universal principles
among systems which had hitherto been considered very
different from each other.

Reflecting the development of the modern theory of critical
phenomena, the rise of physics of complex
networks has been driven by the simultaneous occurrence
of detailed empirical studies of extremely large networks
that were made possible by the advent of affordable high-power
computing and the development of statistical
mechanics tools to analyze the new network models. Prior
to these developments, the networks that were looked at by
physicists belonged to either the class of (i) regular
networks, defined on geometrical lattices, where each node
interacted with all the neighboring nodes belonging to a
specified neighborhood, or (ii) random networks, where any
pair of nodes had a fixed probability of being linked, i.e.,
interacting with each other. The first work that focused
public attention on the new network approach presented a
class of network models that were neither regular nor
random, but exhibited properties of both~\cite{WS98}. 
Such {\em small world networks}, as they were referred to, exhibited high
clustering (with nodes sharing a common neighbor having
a higher probability of being connected to each other than
to other nodes) and a very low average path
length (where the path length between any two nodes is
defined as the shortest number of connected nodes one has
to go through in order to reach one node starting from the
other). As the former property characterized a regular network, while
the latter was typical for a random network, this new class
of networks was somehow intermediate between the
extremes of the two well-known network models, which
was manifest in their construction procedure (Fig. 1).
Several networks occurring in reality, in particular, the
power grid, the actor collaboration network and the neural
connection patterns of the {\em C. elegans} worm, were shown to
have the small-world property. Later, other examples have
been added to this list, including the network of co-active
functional brain areas~\cite{AS06} and the Indian railway system~\cite{SD03}.

\begin{figure}[tbp]
\centering
\includegraphics[width=0.3\linewidth,clip]{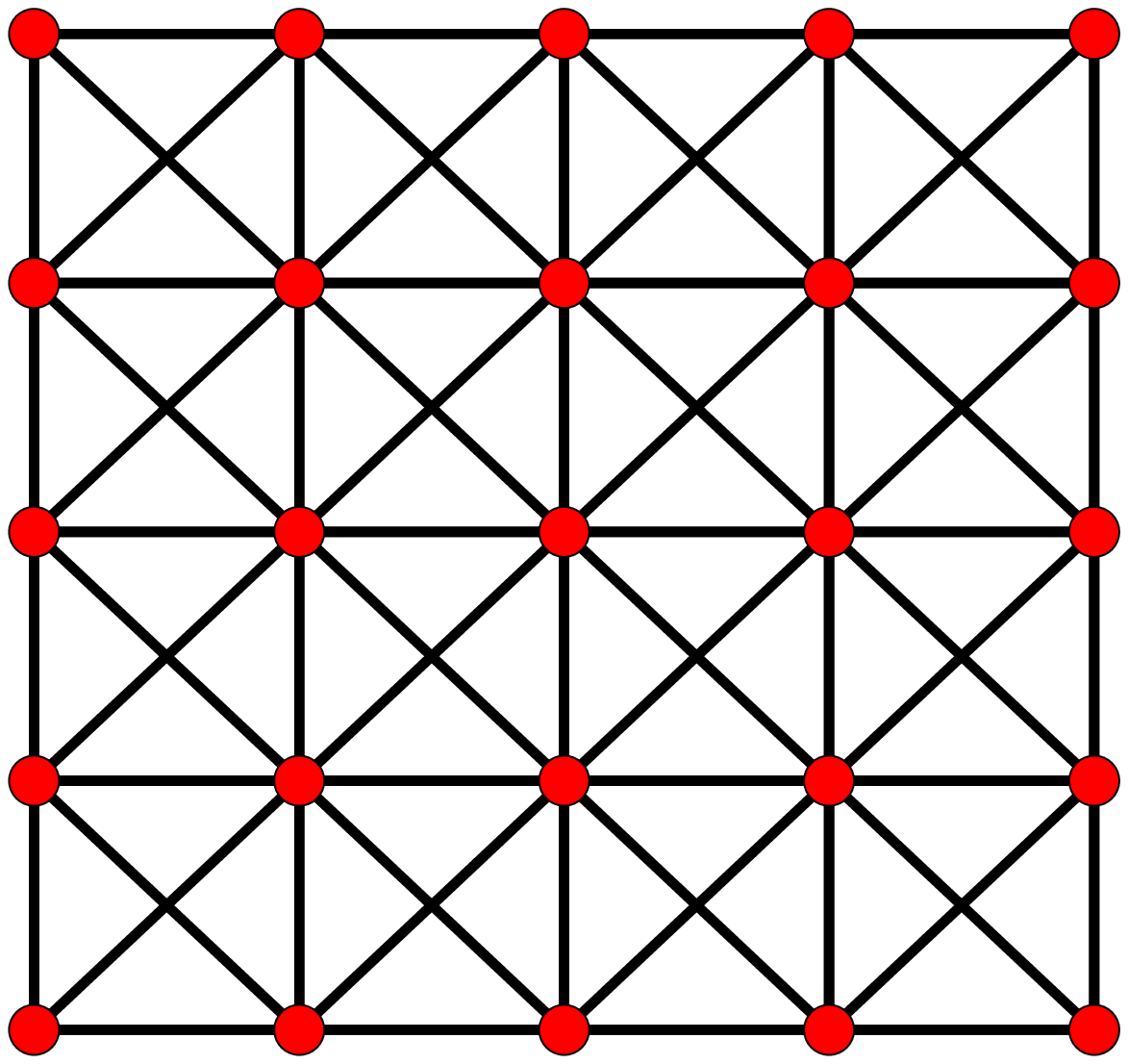}
\includegraphics[width=0.3\linewidth,clip]{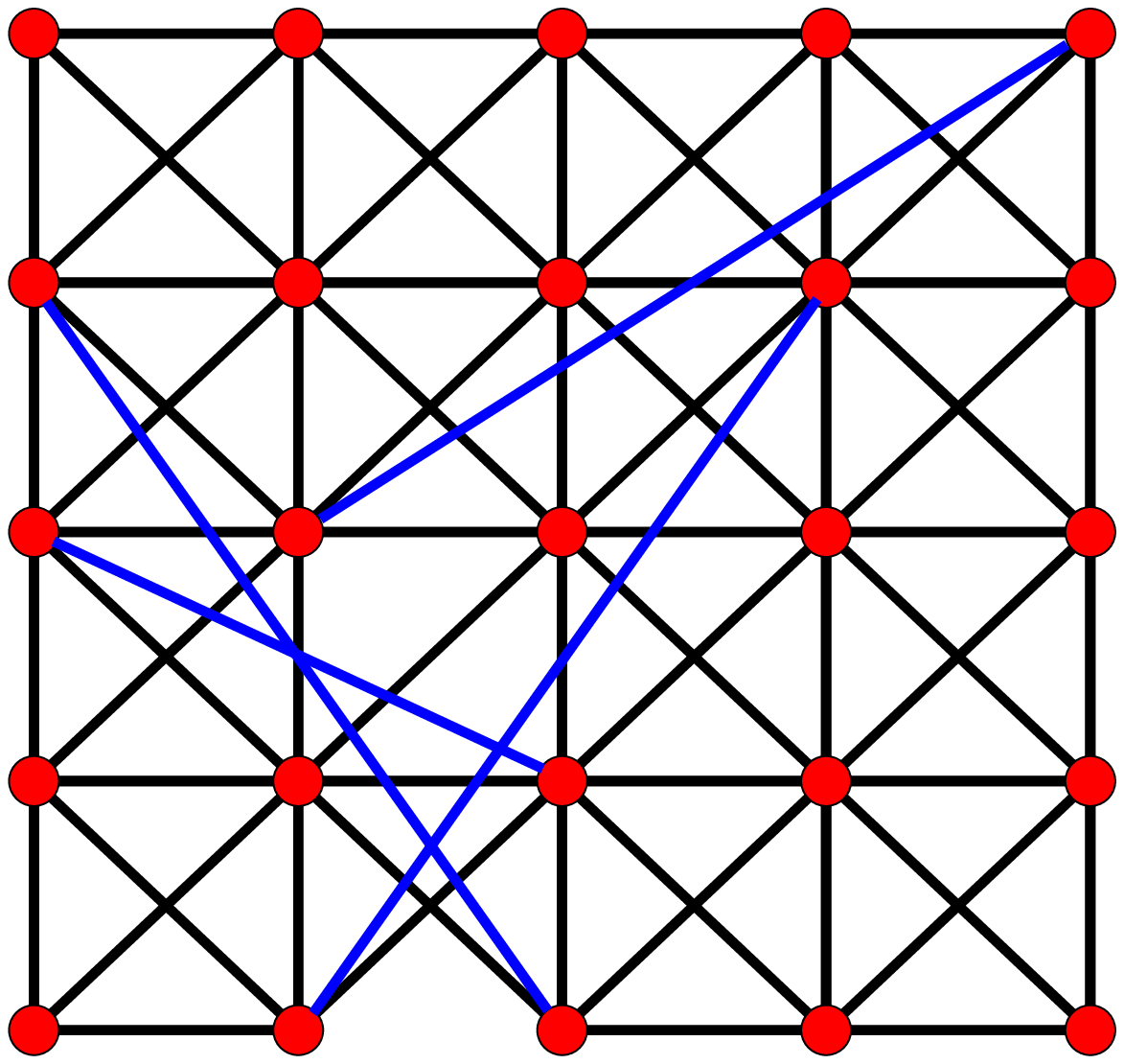}
\includegraphics[width=0.3\linewidth,clip]{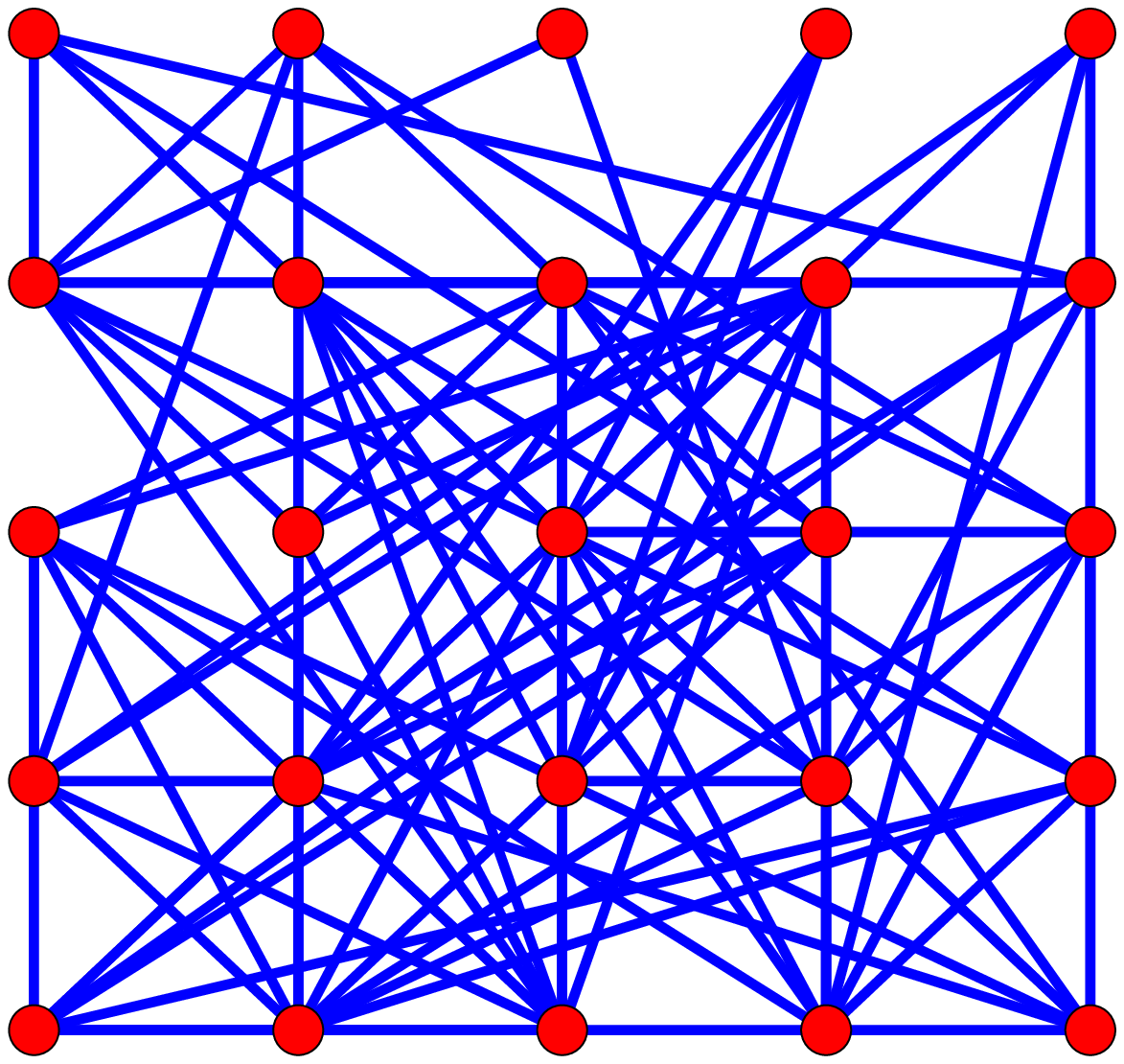}\\
\caption{Constructing a small-world network on a 2-dimensional 
square lattice substrate. Starting from a regular
network (left) where each node is connected to its nearest
and next-nearest neighbors, a fraction $p$ of the links are
rewired amongst randomly chosen pairs of nodes. When all
the links are rewired, i.e., $p = 1$, the system is identical to a
random network (right). For small $p$, the resulting network
(center) still retains the local properties of the regular
network (e.g., high clustering), while exhibiting global
properties of a random network (e.g., short average path
length).}
\label{ssfig1}
\end{figure}
Very soon afterwards, it was discovered that the frequency
distribution of node degree (i.e., the number of links a
node has) exhibits a power-law scaling form for a large
variety of systems including the world wide web~\cite{AB99}. This
further underlined the fact that most networks occurring in
reality are neither regular (in which case the degree
distribution would be close to a delta function) nor random
(which has a Poisson degree distribution), as for both cases
the probability of having a node with large degree (i.e., a hub) would be
significantly smaller than that indicated by the power law
tail of empirically obtained degree distributions. In
addition, it was observed that there exist non-trivial degree
correlations among linked pairs of nodes. For example, a
network where nodes with high degree tend to
preferentially connect with other high degree nodes,
is said to show assortative mixing~\cite{Newman02}. On the other
hand, in a disassortative network, nodes with large number
of links prefer to connect with nodes having low degree.
Empirical studies indicate that most
biological and technological networks are disassortative,
while social networks tend to be assortative~\cite{Newman03}. As
assortative mixing promotes percolation and makes a
network more robust to vertex removal, it maybe hard to
understand why natural evolution in the biological world
has favored disassortativity. However, in a recent study, we
have shown that when one considers the stability of
dynamical states of a network, disassortative networks
would tend to be more robust, and this may be one of the
reasons why they are preferred~\cite{BS05}.

This brings us to the thrust of recent work in the area of
complex networks which has shifted from the initial focus
on purely structural aspects of the connection topology, to
the role such features play in determining the dynamical
processes defined on a network\cite{ST01}. 
Over the past few years, much effort has been made to understand
not only how structure affects
dynamics, and hence function, in a network, but also the
reverse problem of how functional criteria, such as the need
for dynamical stability, can constrain the topological
properties of a network. In this article, some of the principal results
obtained by our group will be briefly described. 
The goal of our research program is to understand the
evolution of robust yet complex biological structures, viz., networks
occurring in reality that are stable against perturbations and,
yet, which can adapt to a changing environment.

%
%
%

\section{Biological Networks: Some examples across length scales}
Before describing our results that are applicable to a 
wide range of networks, we provide motivation for our general approach 
by briefly discussing in this section
a few examples of biological networks. Although they span an enormous 
range of length scales, from $\sim 10^{-8}$ m in the case of protein
contact networks to $\sim 10^5$ m in the case of ecological interaction
networks, they are often subject to similar constraints and may share common 
structural and dynamical properties. Questions asked about networks
in one domain, may often have answers and ramifications in another domain.

\vspace{0.25cm}
\noindent
{\bf Molecular scale: Protein Contact Network.}
\begin{figure}[tbp]
\centering
\includegraphics[width=0.5\linewidth,clip]{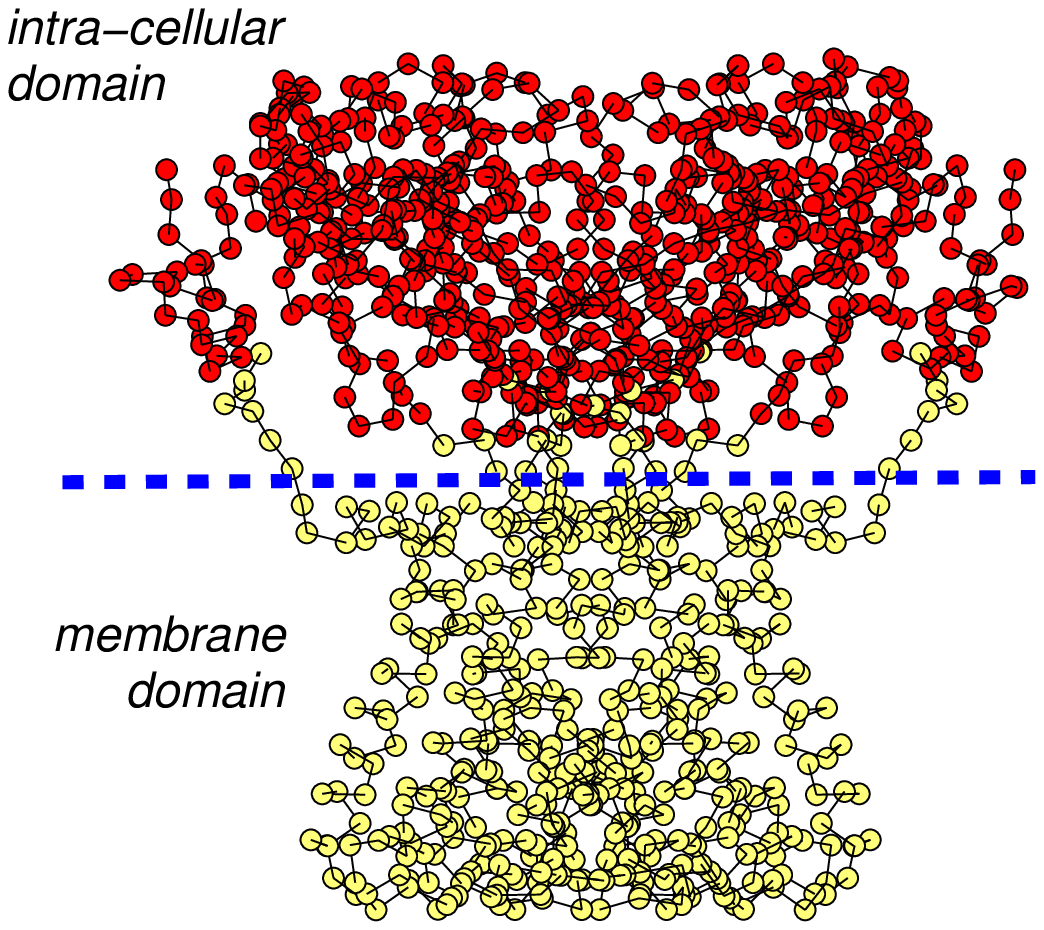}
\includegraphics[width=0.48\linewidth,clip]{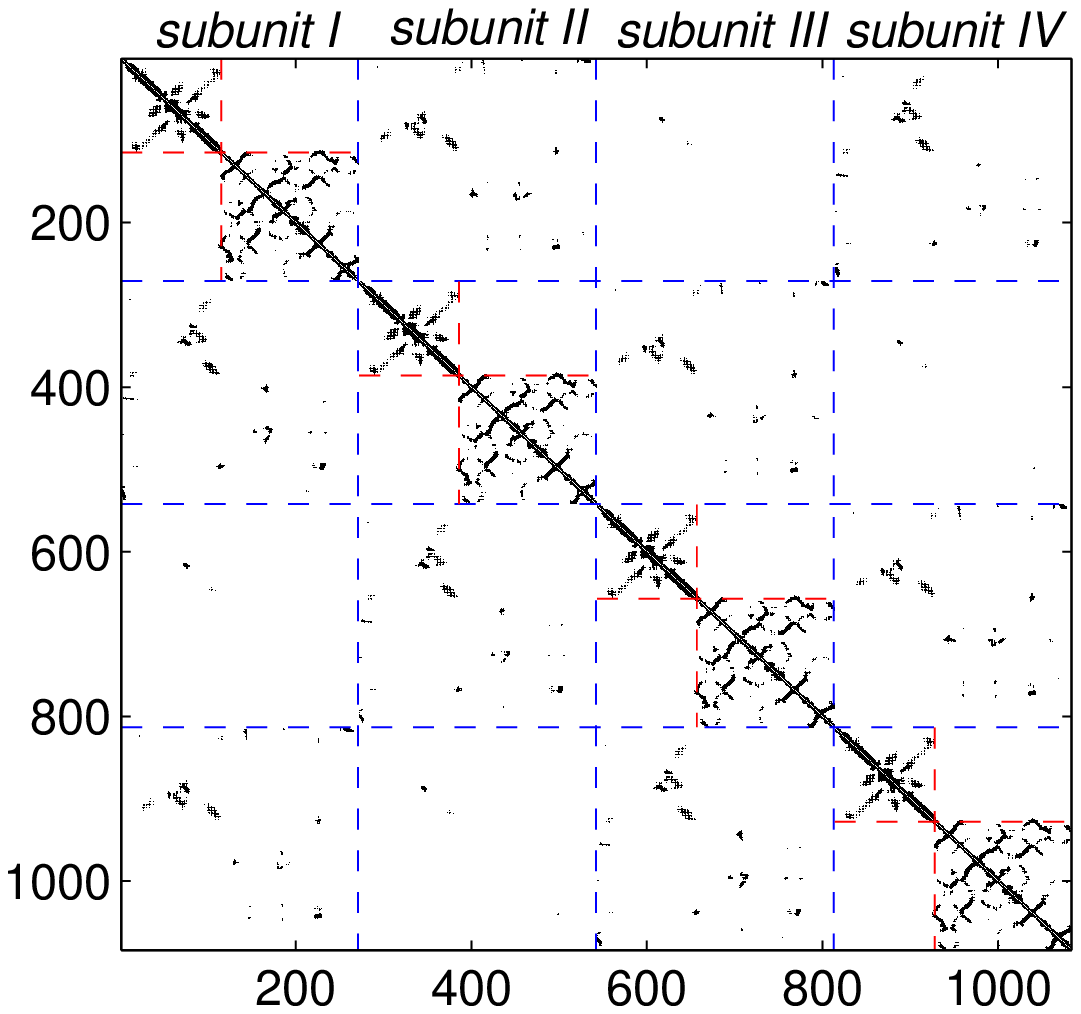}
\caption{Structure of the Kirbac1.1 protein (left) which comprises of 
four identical subunits spanning
the membrane and intra-cellular regions~\cite{Kuo03}. 
The contact network (PCN) is constructed
by considering a cutoff distance of $d_c = 12 \AA$, whose adjacency matrix
is shown for the entire network (right). Each of the four
blocks corresponding to a sub-unit shows
a clear partition into membrane and intra-cellular
compartments, indicating a modular structure. 
}
\label{ssfig:PCN}
\end{figure} 
Protein structure, viewed as a network of non-covalent connections 
between the constituent amino acids, is one of the smallest length scale 
networks in the natural world. Its nodes are the
C$^{\alpha}$ atoms of each amino acid, and their interaction strength is
determined by their proximity to each other. Two nodes are considered 
to be linked if the Euclidean distance between them (in 
3-dimensional space) is less than a cutoff value $d_c$, usually 
between 8-14 $\AA$, 
which is the relevant distance for non-covalent interactions.
Fig~\ref{ssfig:PCN} shows the Kirbac1.1 protein, belonging to the family 
of potassium ion channels involved in transmission of inward rectifying 
current across a cellular membrane~\cite{Kuo03}. It consists of four identical 
sub-units spanning the membrane and intra-cellular regions. The corresponding
protein contact network (PCN) manifests the existence of the identical 
sub-units in the
approximately block diagonal structure of the adjacency matrix. 
In addition, each of these
four blocks can be divided into two modules, corresponding approximately
to the membrane and intra-cellular regions. 

It is easy to see that the PCN shares the features of a small world
network, with the majority of connections being between spatially neighboring
nodes, although there are a few long-range connections. This small-world
property of PCNs for different protein molecules have indeed been
noted several times in the literature (see, e.g., Ref.~\cite{AK07}).
This is probably not very surprising, given that it is also true for 
a randomly folded polymer.
However, in addition, the PCN adjacency network shows a modular structure,
with a majority of connections occurring between nodes belonging to the
same module. This is a feature not seen in conventional models of
small world networks (e.g., the Watts-Strogatz model~\cite{WS98}). 
It is all the more intriguing as we have recently shown that modular 
networks (whatever the connection topology of
individual modules) exhibit the small world properties of high clustering
and low average path length~\cite{PS08}. To identify whether the existence
of modules indeed has a significant effect on protein dynamics
(e.g., during folding),
we look at
the spectral properties of the Laplacian matrix\footnote{The Laplacian is 
also referred to as the Kirchoff matrix (e.g., see Ref.~\cite{HBE97}).} 
${\bf L}$, defined as 
$L_{ii}=k_{i}$, the degree of node $i$, $L_{ij}=-1$ if nodes $i$ and $j$
are connected, $0$ otherwise. The eigenvector for the smallest eigenvalue 
($= 0$), $c^{(1)}$, corresponds to the time-invariant properties of the
system, and has uniform contribution from all components.
The next few smallest eigenvalues dominate the time-dependent
behavior of the protein and these show a relatively 
large spectral gap with the bulk of the eigenvalue spectra. This indicates
the existence of very distinct timescales in the protein dynamics
which approximately correspond to the inter- and intra-modular modes
of motion. As we shall see below, the occurrence of modular structures 
in complex networks and their effect on dynamics
is not just confined to PCNs but appears in many other  biological networks.

\vspace{0.25cm}
\noindent
{\bf Intra-cellular scale: Signaling network.}
\begin{figure}[tbp]
\centering
\includegraphics[width=0.95\linewidth,clip]{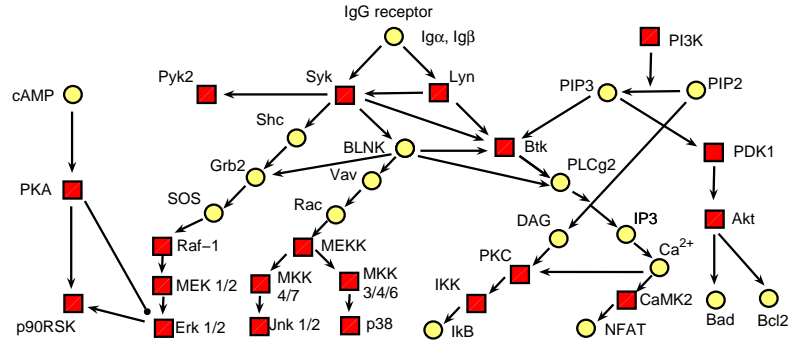}
\caption{A subset of the signal transduction network of the B-cell
antigen receptor (BCR)~\cite{Dhiraj07}. The kinases are represented
by squares, while other molecules (such as, second messengers and adapters) 
are depicted as circles.
}
\label{ssfig:BCR}
\end{figure}
Signal transduction pathways, the process through which a cell responds 
appropriately to a signal or stimulus, involve ordered sequences of 
biochemical reactions carried out by enzymes inside the cell. One
of the most commonly observed class of enzymes in intra-cellular signaling
is that of kinases, which activate target molecules (usually proteins) by
transferring phosphate groups from energy donor molecules such as ATP to
the targets.
This process of phosphorylation is mirrored by the reverse process of
deactivation by phosphatases through dephosphorylation.
Such reaction cascades are activated by second messengers (e.g., cyclic AMP 
or calcium ions) and may last for a few minutes, with the number
of kinase proteins and other molecules involved in the  process 
increasing with every reaction step away from the initial stimulation.
Thus, such a signalling cascade can result in a large response for
a relatively low-amplitude signal. 

Research over the past decade has, however, shown that the classical picture 
of almost isolated cascades linking an unique signal to a specific response
does not explain many experimental results. The adaptability of
intra-cellular signalling is now thought to be a result of 
multiple signaling pathways interacting with one another to form complex 
networks. In this picture, complexity arises from the 
large number of components, many of which have partially overlapping functions,
the large number of links (through enzymatic reactions) among components 
and from the spatial relationship between the components~\cite{WB99}.
Fig.~\ref{ssfig:BCR} shows a small fraction of the signalling network 
downstream of the B-cell antigen receptor (BCR) involved in immune response.
As the breakdown of communication in this network can lead to disease
(a fact that may be utilised by infectious agents for proliferation), 
it is of obvious importance to understand the mechanisms by which the network 
allows the cell response to be sensitive to different stimuli and yet
robust in the presence of intra-cellular noise. With this in mind,
the time evolution of the activity (i.e., phosphorylation) of about 
20 signalling molecules in this network  were recorded in a recent
experiment by Kumar {\em et al}~\cite{Dhiraj07}. Apart from observing
the activation profiles under normal conditions, the network was also 
subjected to a series of perturbations, by serially blocking each 
of these molecules from activating any of the other molecules in the
network. The resulting experimental data, capturing the behavior of
these molecules under 21 different conditions, enabled the detection
of correlations between the activity of these molecules. This showed
that the existing picture of interactions (Fig.~\ref{ssfig:BCR}) is grossly
inadequate in explaining these correlations, e.g., the fact that p38 kinase
seems to influence the activation of a majority of the other molecules,
although it occurs at the end of particular pathway. The results suggest that
the signalling network is in fact a far more densely connected system
than had been previously suspected. It also raises the question of
how certain signals can elicit very specific responses, without 
significant risk of cross-talk between interacting pathways. This brings
us to the issue of whether functional modules can exist in networks,
such that by using positive and negative interactions one can
channel information from the stimulus to the response along specific 
subnetworks only.
 
\vspace{0.25cm}
\noindent
{\bf Inter-cellular scale: Neuronal Network.} 
The above question is of importance not only for information
processing within a cell, but also between cells. The most important example
of the latter process is, of course, the networks of neurons occurring 
in the brain. 
As the nervous system of the nematode {\em C. elegans} comprising 302 neurons
has been 
completely mapped out (in terms of the positions of the neurons, as
well as all their interconnections), it provides a model system 
for studying these issues. We have recently analysed the connection 
topology of the non-pharyngeal portion of the nervous system to which the
majority of the neurons ($\simeq 280$) belong~\cite{CS07}. One of the 
striking observations is that many of the sensory neurons belonging
to different modalities, viz. chemosensation, mechanosensation, etc.,
send signals to the same set of densely connected interneurons which forms
the innermost core of the nervous system. Subsequently, signals are sent from
these interneurons to specific motor neurons which generate appropriate 
muscle response, e.g. moving along a chemical gradient, egg-laying. etc.
It is vital that the signals coming from different sensory
neurons to the same interneurons should not interfere with each other,
as it may result in activating the incorrect motor response. A 
preliminary investigation of a dynamical model for the neuronal network
shows that, a complex set of excitatory and inhibitory links between 
the inter-neurons manages to achieve segregation of the
different functional circuits. This means that, e.g., a mechanical tap signal 
will not elicit egg-laying, even if the tap withdrawal circuit may share
many common interneurons with the egg-laying circuit. Even
more interesting is the fact that such functional modules do not
need the existence of structural modules in the underlying networks. It 
underscores the importance of looking at the nature of the interactions,
which can create complicated control mechanisms to prevent cross-talk 
and enable robust response in the presence of environmental noise.

\vspace{0.25cm}
\noindent
{\bf Inter-organism scale: Epidemic Propagation Network.}
At the scale of individual organisms, such as human beings, one of the
most widely studied networks is that which leads to propagation of epidemics.
The ubiquity of small-world networks in nature implies that some of
the classic theories of epidemiological transmission, based on assumptions
of random connections, may need to be reviewed. In particular, the global
spread of diseases like SARS shows that even a few long-range links can
drastically enhance the propagation of epidemics~\cite{DM07}. This has
led to a series of studies of different disease propagation models
on Watts-Strogatz (WS) or related network models (e.g., see Ref.~\cite{SK05}).
However, as mentioned above, all the structural features of such
networks are also shared by modular networks, which however have very 
different dynamical properties. We have recently shown that while
WS networks have a continuous range of time-scales, modular networks
exhibit very distinct time-scales that are related to intra-
and inter-modular events~\cite{PS08}. Thus, devising an effective 
strategy to counter
the spread of epidemics will have to take into account a detailed 
knowledge of such structures in the social network of contagious and
susceptible individuals.

\vspace{0.25cm}
\noindent
{\bf Inter-species scale: Food Webs.}
Possibly the largest (in terms of length scale) biological networks on
earth are those of interactions between different species in an ecosystem. 
While general ecological networks consist of all possible links, such 
as cooperation and competition, food webs describe the trophic relations,
i.e., between predator and prey.
It is a directed network where the nodes are the various species, 
with prey connected by arrows to predators, the direction
of the arrow indicating the flow of biomass. The links are usually weighted 
to represent the amount of energy that is transferred. It is in the context
of these networks that questions first arose on the connection between
the structural properties of a network and the stability of its
dynamical behavior (see Sec.~\ref{ss:sec4}). Indeed, one can not
only ask what kind of structures allow complex networks to be stable
against ever-present perturbations, but also, how the requirement to
be robust constrains the kind of structures such networks can evolve.
To stress the universality of the questions asked by physicists about
networks, we note that 
like many other networks, food webs also have been shown to have 
modular structure, with
species in each module interacting between themselves strongly 
and only weakly with other species~\cite{Krause03}. As in the other
systems discussed earlier, the role that modularity plays in stabilizing
the dynamics of ecosystems can be seen as a specific instance of a much more 
general question.

Having discussed a few instances of how universal principles about networks
can appear by investigating very different systems in the biological 
world, we now describe certain results of our studies on general network
models.
However, we stress that each of these results have relevance to problems
appearing in the context of specific biological systems.
\vspace{-0.2cm}
\section{From Structure to Dynamics}
The role that the connection topology of a network plays in
the nature of its dynamics has been extensively investigated for
spin models occurring in physics. In fact, such systems had been explored 
for a
long time prior to the recent interest in complex networks,
and many results are known regarding ordering transition in
both regular as well as random structures. More recently, it has been shown
that, for partial random rewiring in a system of sufficiently
large size, any finite value of $p$ (the rewiring probability) 
causes a transition to the
small-world regime, with the Ising model defined on such a
network exhibiting a finite temperature ferromagnetic phase
transition~\cite{BW00}. However, spin models are extremely
restricted in their dynamical repertoire and therefore,
researchers have looked at the effect of introducing other
kinds of node dynamics in such network structures, e.g.,
oscillators. Motivated by recent observations that the brain
may have a connection structure with small-world
properties (see e.g., Ref.~\cite{AS06}), we have examined the effect of
long range connections (i.e., non-local diffusion) over an
otherwise regular network of nodes with links between
nearest neighbors on a square lattice~\cite{SSK07}. The dynamics
considered is that of the excitable type, with the variable
having a single stable state and a threshold. If a perturbation
causes the system variable to exceed the threshold, we see a
rapid transition to a metastable excited state followed by a
slow recovery phase when the system gradually converges
to the stable state. As a result of coupling the dynamics of
individual nodes through diffusive coupling, various spatial
patterns (which may be temporally varying) are observed.
Such dynamics is commonly observed in a large variety of
biological cells such as neurons and cardiac myocytes, as
well as in nonlinear chemical systems such the Belusov-
Zhabotinsky reaction.

\begin{figure}[tbp]
\centering
\includegraphics[width=0.9\linewidth,clip]{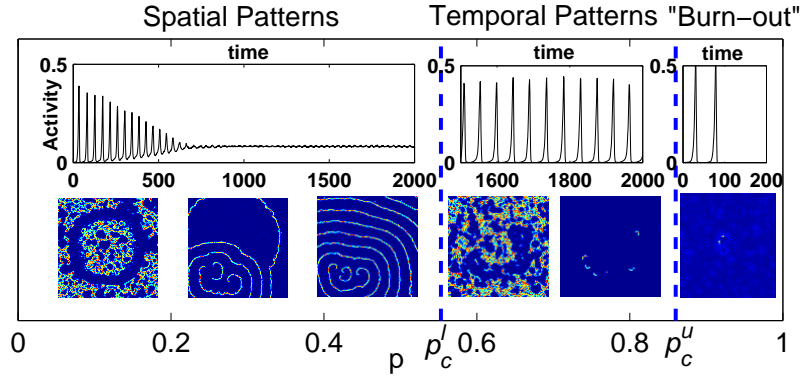}
\caption{Schematic diagram indicating the different dynamical regimes
in a two-dimensional ``small-world'' excitable medium as a function 
of the rewiring probability, $p$. For low $p$, the system exhibits
spatial patterns characterized by single or multiple spirals. At
$p = p_c^l$, there is transition to a state dominated by
temporally periodic patterns that are spatially relatively homogeneous.
Above $p = p_c^u$, all activity ceases after a brief transient.
}
\label{ssfig:kaski}
\end{figure}
In our simulations, by varying the probability of long-range
connections, $p$, we have observed three categories of
patterns. For $0 < p < p_c^l$, after an initial transient period
where multiple coexisting circular waves are observed, the
system is eventually spanned by a single or multiple
rotating spiral waves whose temporal behavior is
characterized by a flat power spectral density. At 
$p = p_c^l$,
the system undergoes a transition from a regime with
temporally irregular, spatial patterns to one with spatially
homogeneous, temporally periodic patterns (Fig.~\ref{ssfig:kaski}). The latter
behavior occurs over the range $p_c^l < p < p_c^u$ 
as a result of
the increased number of long range connections, whereby a
large fraction of the system gets synchronously active and
subsequently goes into the recovery phase. Beyond the
upper critical value $p_c^u$, there is no longer any self-sustained
activity in the system as all nodes converge to the stable
state. The patterns in each regime were found to be
extremely robust against even large perturbations or
disorder in the system.

Our model explains several hitherto unexplained
observations in experimental systems where non-local
diffusion had been implemented~\cite{STS06}.
In addition, by identifying
the long-range connections with those made by neurons and
the regular network with that formed by the glial cells in the
brain, our results provide a possible explanation of why
evolution may have preferred to increase the number of
glial cells over neurons (with a ratio of more than 10:1 for
certain parts of the human brain) in order to maintain robust
dynamical patterns as brain size increased. It also points
towards possible functional role of small-world brain
topology in the occurrence of dynamical diseases such as
epileptic seizures and bursts. More generally, our work
shows how non-standard network topologies can influence
system dynamics by generating different kinds of
spatiotemporal patterns depending on the extent of nonlocal
diffusion.

\vspace{-0.2cm}
\section{From Dynamics to Structure}
\label{ss:sec4}
An important functional criterion for most networks
occurring in nature and society is the stability of their
dynamical states. While earlier studies have concentrated
on the robustness of the network when subjected to
structural perturbations (e.g., removal of node or link), we
have looked at the effect of perturbations given to the
steady states of network dynamics. In particular, the
question we ask is whether networks become more
susceptible to small perturbations as their size (i.e., number
of nodes $N$) increases, the connections between the nodes
become denser (i.e., increased connectance probability $C$)
and the average strength of interaction ($s$) increases. This is
related to a decades-old controversy, often referred to as the
stability-complexity debate. In the early 1970s, May~\cite{May73} 
had shown that for a model ecological network, where
species are assumed to interact with a randomly chosen
subset of all other species, an arbitrarily chosen equilibrium state
of the system becomes unstable if any of the parameters
determining the network's complexity (e.g., $N$, $C$ or $s$) is
increased. In fact, by using certain results of random matrix
theory, the critical condition for the stability of the network
was shown to be $N C s^2 < 1$ (May-Wigner theorem)~\cite{May73}. This
flew against common wisdom, gleaned from large number
of empirical studies as well as naive reasoning, which
dictated that increased diversity and/or stronger interactions
between species results in more robust ecosystems. Thus,
ever since the publication of these results, there
have been attempts to understand the reason behind the
apparent paradox, especially as this result
relates not only to ecological systems but extends to all
dynamical networks for which the stability of equilibria
have functional significance, e.g., in intra-cellular biochemical
networks where the concentrations of different molecules
need to be maintained within physiological levels. Two of
the common charges leveled against the theoretical model of May
is that (i) it assumes the interaction network to be random
whereas naturally occurring networks may have
certain kinds of structures, and (ii) the linear stability analysis
assumes the existence of simple steady states (viz., fixed
point attractors), which may
not be the case for real systems that may either be having
oscillations or be in a chaotic state.

\begin{figure}[tbp]
\centering
\includegraphics[width=0.48\linewidth,clip]{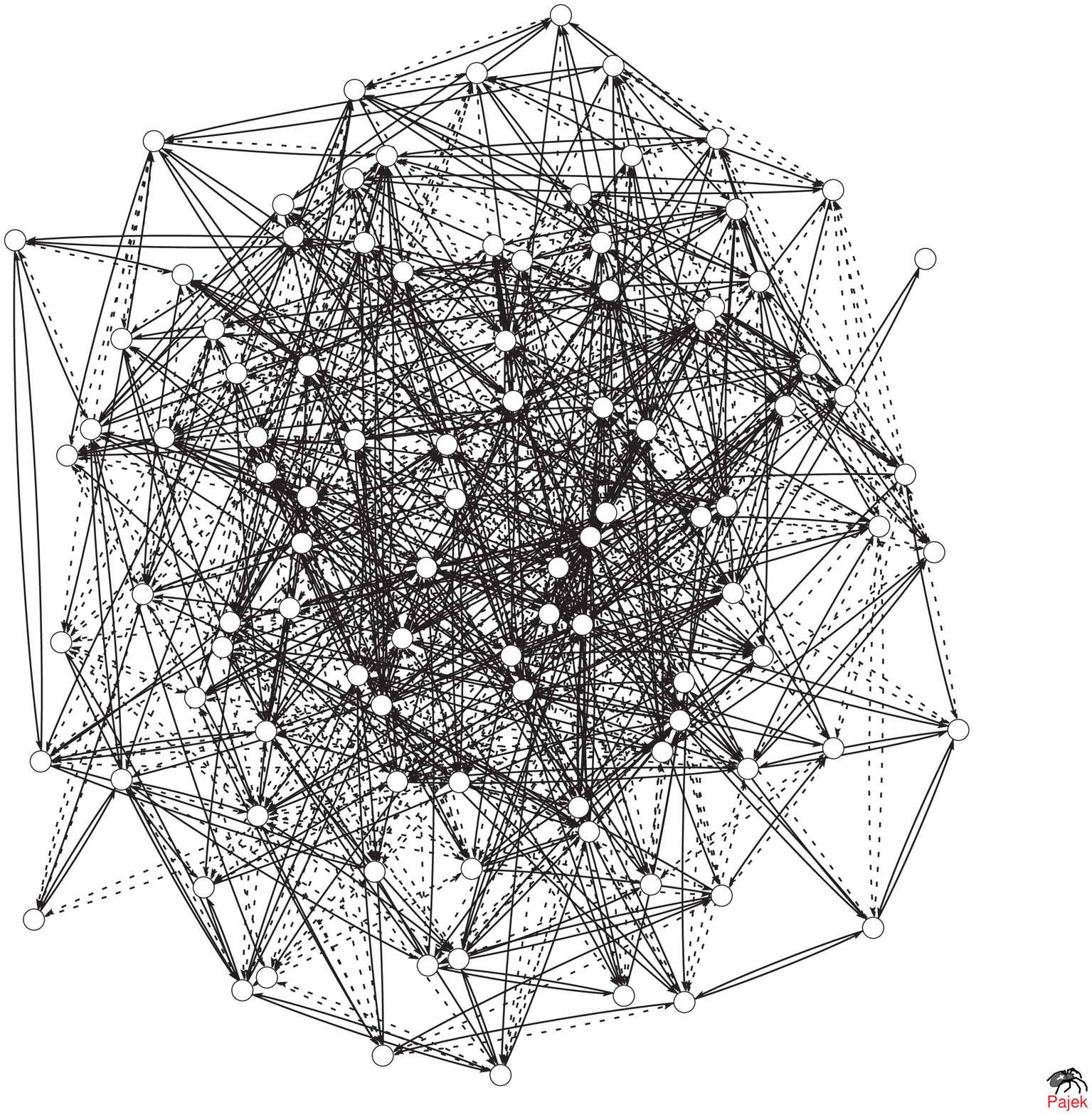}
\includegraphics[width=0.48\linewidth,clip]{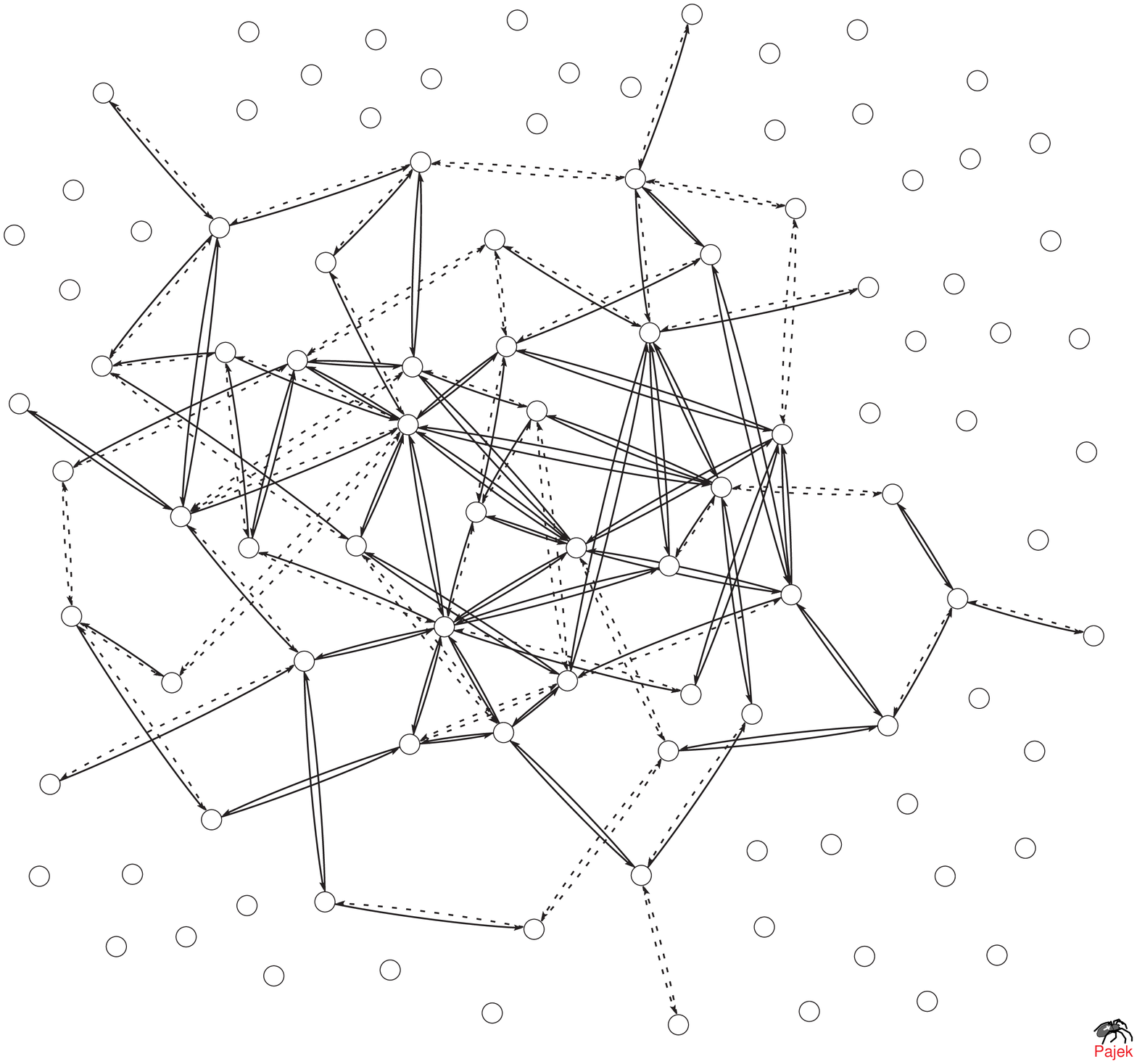}\\
\caption{Evolution of a network with non-trivial dynamics at
the nodes. The initial (left) and final asymptotic (right)
networks are shown. Only nodes having persistent activity
are connected to the network. The figures were drawn using
Pajek software.
}
\label{ssfig:pajek}
\end{figure}
In our work on dynamical systems defined on networks, we
have tried to address both of these lines of criticism (see Ref.~\cite{WI07}
for a recent discussion of our results from the perspective of ecosystem
robustness). For
example, focusing on the question of inadequacy of linear
stability analysis, we have considered networks with nontrivial
dynamics at the nodes, spanning the range from
simple steady states to periodic oscillation and fully
developed chaos, and measured the robustness of the
dynamics with respect to variations in $N$, $C$ and $s$~\cite{SS05,SS06}.

Each node in our model network has a dynamical variable
associated with it, which evolves according to a well-known
class of difference equations commonly used for
modeling population dynamics. By varying a nonlinear
parameter, the nature of the dynamics (i.e., whether it
converges to a steady state or undergoes chaotic
fluctuations) at each node can be controlled. However, in
the absence of coupling, each node will always have a
finite, positive value for its dynamical variable. When
coupled in a network (initially in a random fashion), with
links that can have either positive or negative weights, it is
possible that as a result of dynamical fluctuations, the
variable for some nodes can become negative or zero. As
this implies the absence of any activity, the
corresponding node is considered to be ``extinct'' and thus
isolated from the network. This procedure may create
further fluctuations and cause more nodes to get ``extinct'',
resulting in gradual reduction of the size of the network
(Fig.~\ref{ssfig:pajek}). The final asymptotic size of the network, relative
to its initial size, is a measure of its robustness - the more
robust network being one with a higher fraction of nodes
having persistent activity.
Analysis showed that the network robustness (as measured
by the above global criterion) not only decreased with $N$, $C$
and $s$, as expected from local stability analysis, but
actually matched the May-Wigner theorem quantitatively~\cite{SS05}.
In addition, the asymptotic network exhibited robust
macroscopic features: (a) the number of persistently active
nodes was independent of the initial network size, and (b)
the asymptotic number of links between these persistently
active nodes was independent of both the initial size and
connectivity~\cite{SS06}. This is all the more surprising as the
removal of nodes (and hence, links) is not guided by any
explicit fitness criterion but rather emerges naturally from
the nodal dynamics through fluctuations of individual node
properties. Our results imply that asymptotically active
networks are non-extensive: when two networks of size $N$
are coupled to each other (with the same connectance as the
individual networks), although the resulting network
initially has a size $2N$, the ensuing dynamical fluctuations
will reduce its size to $N$. This implies that simply increasing
the number of redundant elements is not a good strategy for
designing robust systems.

We have also looked at the effect of empirically reported
structures, such as small-world connection topology and
scale-free degree distribution, on the dynamical stability of
networks. Our results indicate that, in general, introducing
such structural features do not alter the outcome expected
from the May-Wigner theorem~\cite{BS05,SI05}. However, these
details can indeed affect the nature of the stability-instability
transition, for example, the transition exhibiting a
cross-over from being very sharp (resembling first-order
phase transition) for a random network to a more gradual
change as the network becomes more regular in the small world
regime~\cite{SI05}. 

\vspace{-0.2cm}
\section{Evolution of Robust Networks}
This brings us to the issue of how complex networks can be
stable at all, given that the May-Wigner theorem seems to
hold even for networks that have structures similar to those
seen in reality and where non-trivial dynamical situations
have also been considered. The solution to this apparent
paradox lies in the observation that most networks that we
see around us did not occur fully formed but emerged
through a process of gradual evolution, where stability with
respect to dynamical fluctuations is likely to be one of the
key criteria for survival. In earlier work, we have shown
that a simple model where nodes are gradually added to or
removed from a network according to whether this results
in a dynamically stable network or not, results in
a non-equilibrium steady state where the network is
extremely robust~\cite{WSB02}. The robustness is manifested by
increased resistance and resilience, as well as, decreased
probability of large extinction cascades, when the network
size (i.e., the system diversity) is increased. Thus, our
results reconcile the apparently contradictory conclusions
of the May-Wigner theorem and a large number of
empirical studies.

\begin{figure}[tbp]
\centering
\includegraphics[width=0.95\linewidth,clip]{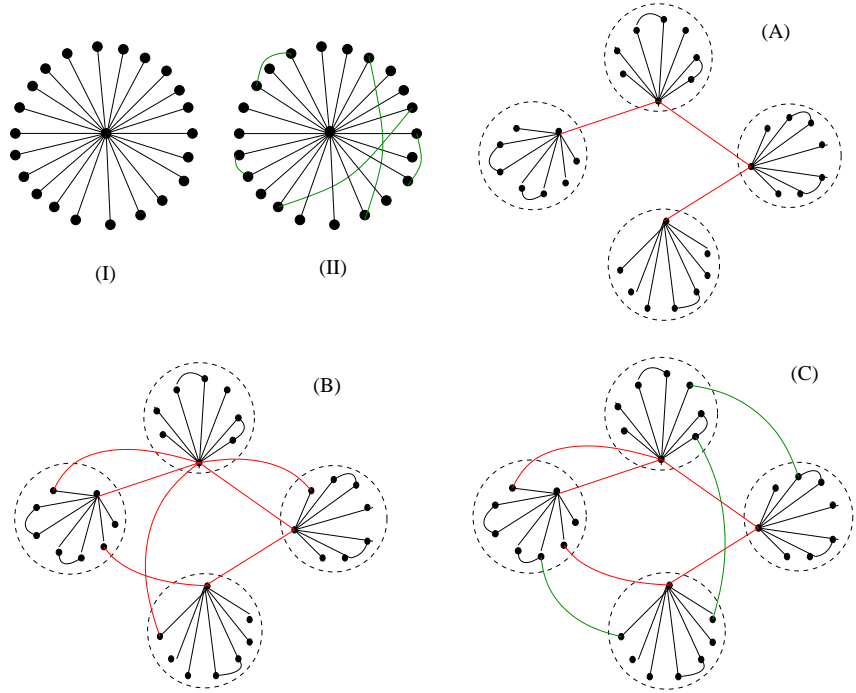}\\
\caption{Networks with (I) star and 
(II) clustered star connection topology, can form the fundamental building
blocks of different types of modular networks. Network configurations having 
clustered star modules can be constructed by (A) connecting different modules
by single undirected links among the hub nodes, or (B) connecting
nodes of a module to another module only through the hub node of the latter,
or (C) connecting nodes of a module randomly to any node of another module.
}
\label{ssfig:robust1}
\end{figure}

More recently, we have shown that model networks can
evolve many of the observed structural features seen among
networks in the natural world, by taking into account the
fact that most such systems have to optimize between several
(often conflicting) constraints, which may be structural as well as 
dynamical in nature. In
particular, most networks need to have high communication
efficiency (i.e., low average path length) and 
low connectivity (to reduce the resource
cost involved in maintaining many links) while being
stable with respect to dynamical perturbations. 
If a network satisfied only the first two constraints,
the optimal structure would have been that of a star 
(Fig.~\ref{ssfig:robust1}). Even if the resource cost constraint is somewhat
relaxed, so that the network can have more links than the minimum necessary
to make it connected, the resulting optimal configuration is slightly
modified to that of a ``clustered'' star. However, we note that the
dynamical equilibria in such systems
would be extremely unstable with respect to small perturbations.
This is because the rate of growth of small perturbations is related to
the maximum degree of the network, which, in the case of a star or a clustered
star, is almost identical to the system size. It is easy to see that
dividing the network into multiple stars, connected to each other,
will reduce the maximum degree and hence increase the stability.
Indeed, our results
show that simultaneous optimization of all three constraints
result in networks with modular structure, i.e., subnetworks
with a high density of connections within themselves
compared to between distinct subnetworks, where each module possesses
a prominent hub~\cite{PS07} (see
Fig.~\ref{ssfig:robust1} for possible configurations of such
modular networks). As these
evolved systems also exhibit heterogeneous degree
distribution, our findings have implications for a
wide range of systems in the biological and technological
worlds where such features have been observed.

\vspace{0.25cm}
\noindent
{\small
{\bf Acknowledgements}: I would like to thank my collaborators with
whom the work described here has been carried out, in particular,
R.K. Pan, S. Sinha, N. Chatterjee, M. Brede, C.C. Wilmers, J. Saram\"{a}ki
and K. Kaski, as well as, S. Vemparala, D. Kumar, K.V.S. Rao and B. Saha 
for helpful discussions.}

%

\bibliography{}



\printindex

\end{document}